\documentclass{article}

\usepackage{microtype}
\usepackage{graphicx}
\usepackage{subfigure}
\usepackage{booktabs} 
\usepackage{amsmath, multirow, array, verbatim, amsfonts}
\usepackage[inline]{enumitem}
\usepackage{enumitem}
\usepackage{caption}

\usepackage{hyperref}

\usepackage[accepted]{icml2019}

\icmltitlerunning{Effectiveness of Voice Conversion on SID and ASR Systems}

\begin{document}

\twocolumn[
\icmltitle{Measuring the Effectiveness of Voice Conversion on\\
           Speaker Identification and Automatic Speech Recognition Systems}



\icmlsetsymbol{equal}{*}

\begin{icmlauthorlist}
\icmlauthor{Gokce Keskin}{intel}
\icmlauthor{Tyler Lee}{intel}
\icmlauthor{Cory Stephenson}{intel}
\icmlauthor{Oguz H. Elibol}{intel}

\end{icmlauthorlist}

\icmlaffiliation{intel}{Intel AI Lab, Santa Clara, CA, USA}
\icmlcorrespondingauthor{Gokce Keskin}{gokce.keskin@intel.com}

\icmlkeywords{Machine Learning, ICML}

\vskip 0.3in]



\printAffiliationsAndNotice{}  

\begin{abstract}
This paper evaluates the effectiveness of a Cycle-GAN based voice converter (VC) on four  speaker identification (SID) systems and an automated speech recognition (ASR) system for various purposes. Audio samples converted by the VC model are classified by the SID systems as the intended target at up to 46\% top-1 accuracy among more than 250 speakers. This encouraging result in imitating the target styles led us to investigate if converted (synthetic) samples can be used to improve ASR training. Unfortunately, adding synthetic data to the ASR training set only marginally improves word and character error rates. Our results indicate that even though VC models can successfully mimic the style of target speakers as measured by SID systems, improving ASR training with synthetic data from VC systems needs further research to establish its efficacy.
\end{abstract}

\section{Introduction}
\label{sec:intro}
Converting the voice of a source speaker to a target style has been studied in the context of voice conversion and speaker de-identification \cite{Stylianou:1998,Jin:2009}. In speaker de-identification, an utterance from a source speaker is converted to a style such that the source speaker is anonymized. Voice conversion (VC) is a more complex task where a source speaker utterance is converted to match the style of a specific target speaker. Both of these conversions preserve the content of the source utterance. 

Recent work on Cycle-GAN based VC has demonstrated reasonable performance both in naturalness of converted speech and the style conversion when evaluated by humans \cite{Kameoka:2018}. The model is trained to perform conversions between a relatively small set of four speakers. This encouraging result opens up questions about the susceptibility of automated systems, such as speaker identification and automated speech recognition, to VC clips. In this paper, we aim to answer two such questions: 
\setlist[itemize]{noitemsep, topsep=0pt}
\begin{itemize}[noitemsep]
\item Do VC systems successfully impersonate the target according to speaker ID systems?
\item If VC systems do perform good impersonation, can we use voice conversion as data augmentation for training ASR systems?
\end{itemize}

In order to answer the first question, we use the Cycle-GAN based VC model in \cite{gkeskin_cgan} that can convert among 291 speakers. Once trained, this model is used to convert source utterances to the style of desired target speakers. Four separate SID systems are used to measure the style transfer quality of the VC. The SID models we use are independently trained on ground-truth (non-converted) samples from the target speakers. Our results show that converted utterances are classified as the intended target (top-1) at up to 46\% accuracy among 291 potential targets, with  top-10 accuracy above 70\% for certain SID systems (Section \ref{ssec:sid_results}).

In order to answer the second question, we investigate if the trained VC model can be used as an augmentation method when training an ASR system. We pick a subset of speakers from Librispeech, a dataset with non-parallel utterances \cite{Panayotov:2015}. We train the Deep Speech 2 ASR model \cite{DeepSpeech2} using utterances from the chosen speakers. We use the VC model to convert the chosen speakers' utterances among each other in order to increase the diversity of the content for each speaker, and re-train the ASR system with converted utterances added to the training set. Despite the fact that style conversions are rated highly by the SID systems, we find that ASR models trained with both real and converted utterances do not have significant improvement of WER and CER in the validation set compared to an ASR model trained with only real utterances (Sec. \ref{ssec:asr_results}). 

The counter-intuitive results in ASR augmentation follows the results observed in \cite{biggan_aug}, where GAN-generated synthetic images are observed to improve the accuracy of an ImageNet classifier only marginally (Sec. \ref{sssec:cv_synth_aug}). Based on the results in these two different domains, we believe that synthetic data augmentation needs further study before being widely adopted in training supervised models. Additionally, studying the contrasting results obtained from SID and ASR systems can serve as a starting point in understanding the problems associated using synthetic data for training ASR systems. 

\section{Prior Work}
\label{sec:prior}
\subsection{GANs and Voice Conversion}
\label{ssec:prior_vc}
Generative adversarial networks proposed by \cite{GoodfellowGAN} have demonstrated the rich capabilities of deep learning by producing high resolution, realistic images conditioned to different characteristics of a given input \cite{styleGAN}. The addition of a \textit{cycle loss} in Cycle-GANs has enabled domain transfer in tasks where parallel data does not exist, such as transforming paintings to photos \cite{Zhu:2017}. These generative models have been successfully used for voice conversion, where utterances of a source speaker are converted to the style of a desired target speaker without the need for text transcriptions or parallel data \cite{Kameoka:2018}. Although other VC methods exist, we use the Cycle-GAN VC in \cite{gkeskin_cgan} since it performs reasonably well in human subjective tests and doesn't require parallel data or text transcripts for voice conversion. Additionally, it can perform conversions from/to speakers that are not in the training set, with only seconds of unlabeled speech needed for a new target speaker.

\subsection{SID Systems for Measuring VC Performance}
\label{ssec:prior_sid}
Most VC models perform conversions among a handful of speakers and measure style conversion quality with A/B or ABX testing on human subjects \cite{Kameoka:2018, Kameoka:2018_2, kameoka_convs2s, tanaka_atts2s,  Saito2018}. Although human subjective testing is valuable, it does not give any information about how SID systems respond to converted samples and it is not scalable for conversions among dozens of speakers. 

A few recent studies use SID systems to evaluate synthesized speech. The VC model in \cite{gkeskin_cgan} uses a single i-vector based SID model to evaluate the effectiveness of style conversions among more than three dozen speakers. An SID system is used in \cite{Jin:2009} to evaluate the quality of voice de-identification, where a source utterance is modified such that it cannot be traced back to the source speaker. In text-to-speech synthesis, a deep learning based SID system has been used to evaluate the output style \cite{deepvoice2}.

In this work, we measure if samples converted by the VC model are classified as the intended target by four separate automated SID systems: Two based on i-vectors and two based on deep learning (Sec. \ref{ssec:sid_models}).  

\subsection{Synthetic Data Augmentation}
\label{ssec:prior_synth_aug}

\subsubsection{In Computer Vision Domain}
\label{sssec:cv_synth_aug}
GAN-generated images have been used during training for a medical image classification model in \cite{gan_liver_lesion}, where real data is augmented with relatively small ($64 \times 64$) synthetic images to improve classification accuracy. Augmentation with larger images ($128 \times 128$) is shown to improve accuracy only in cases with very limited data \cite{bowles18}. No statistically significant improvement is seen when augmentation is performed when more than a few hundred real images are available; in fact, degradation is seen for one of the  datasets. 

Higher resolution ($256 \times 256$) synthetic image augmentation in the more widely-used ImageNet dataset shows degradation in classification accuracy as amount of synthetic data is increased, with marginal improvement seen for low numbers of synthetic images \cite{biggan_aug}. Training with only synthetic data significantly increases error rates.

\subsubsection{In Speech Domain}
\label{sssec:speech_synth_aug}
To the best of our knowledge, GAN-based VC models have not been used for synthetic data augmentation in ASR models. The closest prior work focuses on using short duration, unintelligible speech frames generated by a GAN as an ASR augmentation method \cite{hu_gan_18}, where WER is reduced from 8.84\% to 8.37\%. The Cycle-GAN based VC model in \cite{Hosseini:2018} can transfer speech between male/female domains, rather than specific speakers. This VC is used to convert male speech input to female during  inference in an ASR model that is trained using only female speech data.

Tacotron-based TTS models \cite{tacotron, tacotron2} have been used to generate synthetic speech to improve ASR performance. In \cite{nvidia-tts-aug}, synthetic speech generated by TTS is added in 1:1 ratio to real data, with WER decreasing from 5.10\% to 4.66\%. The improvement degrades when more synthetic speech is added to the training set. When the TTS model is jointly trained with an ASR, CER is reduced from 17.35\% to 9.86\% \cite{speech_chain}. We note that the latter result incorporates additional unlabeled, non-synthetic speech data to the training set and uses semi-supervision to provide this significant gain.

In this work, we train an ASR system on labeled (utterance, text) pairs from a set of speakers and record WER/CER values. Training data is non-parallel, i.e. each speaker utters different texts. We then use our trained VC model to convert utterances of training set speakers to each other, and add (converted utterance, text) pairs to the training set. This improves the diversity of uttered texts for each speaker, and can be seen as an additional augmentation method for ASR training. We measure the effectiveness of VC-based augmentation by training the ASR system with augmented data and compare WER/CER values between the two cases.

\section{Implementation}
\label{sec:implementation}
\subsection{VC Model}
\label{ssec:vc_model}
The VC model in this paper closely follows the Cycle-GAN based method in \cite{gkeskin_cgan}. VC converts the log-magnitude mel-spectrogram of the source utterance to the style of the target speaker. Output mel-spectrogram is rebuilt in the audio domain using Griffin-Lim algorithm \cite{Griffin:1984}. Although audio artifacts introduced by Griffin-Lim reduces the naturalness in \cite{gkeskin_cgan}, human listeners rate the conversion quality reasonably high. In this paper, our main focus is to evaluate the effectiveness of conversions in different automated systems. VC model does not have any information about the SID models described in Sec. \ref{ssec:sid_models}. 

\subsection{SID Models}
\label{ssec:sid_models}
Four SID models are used to measure the effectiveness of the VC model. The first is based on the Voxceleb SID system in \cite{voxceleb} and uses linear spectrograms. The second model follows the speaker discriminative model architecture in \cite{deepvoice2} and uses MFCCs. 

i-vector1 and i-vector2 are based on the recipe described in \cite{kaldiSPID} and uses Kaldi speech recognition toolkit \cite{Povey_ASRU2011}. i-vector1 uses \texttt{$low\_freq$ = 40Hz, $vtln\_low$ = 60Hz} and i-vector2 uses \texttt{$low\_freq$ = 60Hz, $vtln\_low$ = 80Hz}. Other MFCC parameters are kept the same (\texttt{$f_s$ = 16KHz}, \texttt{$frame\_length$ = 25ms}, \texttt{ $frame\_shift$ = 10ms}, \texttt{$high\_freq$ = 7800}, \texttt{$num\_ceps$ = 20},  \texttt{$vtln\_high$ = 7200}). The reason for including two similar SID systems in the evaluation is to see if minor modifications in hyperparameters impact the effectiveness of the VC model. All SID models are trained on the same set of speakers (Sec. \ref{ssec:dataset}).

\subsection{ASR Model}
\label{ssec:asr_model}
Deep Speech 2 is the ASR system used for our experiments \cite{DeepSpeech2}. We perform random tempo, pitch shift, amplitude and noise augmentation to both genuine and converted audio files during training. Background noise data is obtained from the Speech Commands dataset \cite{speech_commands}. ASR model uses linear spectrograms of the input audio files during training.

\subsection{Dataset}
\label{ssec:dataset}
VC training is performed on Librispeech train-clean-100 dataset for 251 speakers, approximately 25 minutes of total utterances per speaker \cite{Panayotov:2015}. All four SID models are trained on a combined dataset of 251 speakers from the train-clean-100 and 40 speakers from dev-clean (291 total). ASR system is trained on 16 randomly picked speakers from train-clean-100. This small subset is chosen to more easily observe the impact of synthetic data augmentation on the ASR model.

\section{Results}
\label{sec:results}

\subsection{Effectiveness of VC on SID Models}
\label{ssec:sid_results}
We evaluate the effectiveness of the VC model on four SID systems described in Sec. \ref{ssec:sid_models}. All four SID models achieve above 98\% top-1 accuracy in the evaluation set held out from the 291 speakers (Sec. \ref{ssec:dataset}). 

Eight source speakers are randomly chosen from train-clean-100 dataset, split evenly between male and female. Sixteen speakers are chosen randomly from the train-clean-100 as targets, split evenly among genders. We convert 10 audio samples from each source speaker to each of the target speakers, and rebuild the converted audio in raw audio domain using the Griffin-Lim algorithm. Converted samples are then input to the SID models and top K guesses of each model is recorded (K $\in \{1, 3, 5, 10, 20\}$). If the intended target of a conversion is in the top K guesses, we record this conversion as a success.

\begin{table}[t!]
\centering
\caption{Classification Accuracy of Converted Samples in Different SID Systems}
\label{tab:sid_results}
\resizebox{0.95\columnwidth}{!}{
\begin{tabular}{c|c|c|c|c|c}
\multicolumn{1}{ c }{ SID } & \multicolumn{5}{| c }{Accuracy (Percent)}\\
\multicolumn{1}{ c }{Model} & \multicolumn{1}{| c |}{Top-1} & Top-3 & Top-5 & Top-10 & Top-20 \\
\toprule[.2em]
 Voxceleb & 46.0 & 62.7 & 69.8 & 77.7 & 84.0\\
 DeepVoice 2 & 33.9 & 47.7 & 54.9 & 66.3 & 77.6\\
 i-vector1 & 13.1 & 24.2 & 32.1 & 42.7 & 56.6\\
 i-vector2 & 11.9 & 23.0 & 29.8 & 42.8 & 57.6\\
 \toprule[.15em]
 Chance & 0.3 & 1.0 & 1.7 & 3.4 & 6.8
\end{tabular}
}
\end{table}

Table \ref{tab:sid_results} shows the average Top-K accuracy of the conversion samples for the three SID models. Top-1 accuracy measures if the SID system classifies the converted audio as coming from the intended target of the conversion among 291 potential speakers. The last row shows the audio being classified as the intended target by an SID model that performs uniform random guesses. Top-1 accuracy in deep learning based Voxceleb SID model reaches 46.0\%, with Deep Voice 2 model lower at 33.9\%. i-vector based models are ``fooled" with lower top-1 accuracies (13.1\% and 11.9\%). Top-20 accuracy is above 50\% for all SID models. Minor differences in the hyperparamaters i-vector1 and i-vector2 impact accuracy by about one percentage point.

The results demonstrate that the VC model achieves good style conversion, as measured by automated SID systems, even when trained to convert between dozens of speakers. This is encouraging given that the VC model uses different spectral features for conversion than the SID models use for classification (Sec. \ref{ssec:vc_model} and \ref{ssec:sid_models}). Additionally, VC model does not have any information about the SID models, and it is \textit{not} trained to perform adversarial attacks on them.

It is interesting to see that i-vector models are ``fooled" at lower rates than the deep learning models. One potential reason might be that deep learning based SID models might have learned to recognize background noise or recording artifacts for each speaker, and although independently trained, VC model learns to mimic these same artifacts. i-vector based SID models might be extracting features that strip out these artifacts, making them less prone to classify synthetic audio as the intended target. Even so, synthetic audio is still classified above 10\% top-1 accuracy by the i-vector SIDs.

\subsection{Using VC to Augment ASR Models}
\label{ssec:asr_results}
Given the high style conversion accuracy of converted samples as rated by the SID models, we investigate if the VC model can be used as an augmentation method for ASR training. We randomly pick 16 speakers from train-clean-100 dataset and train Deep Speech 2 ASR system on ground-truth (utterance, text) pairs as a baseline. Training set size is chosen intentionally small to easily observe any potential impact of VC augmentation, and high values of WER/CER are expected due to this small size.

Utterances of the chosen speakers are converted among each other and added to the training set. Since Librispeech train-clean-100 is non-parallel (speakers utter different texts), converted samples increase the diversity of the (utterance, text) pairs for each speaker. In the presence of a perfect voice converter, one would expect lower WER/CER when the ASR system is trained with the augmented dataset.

\begin{figure}[t!]
\begin{center}
\includegraphics[width=0.7\columnwidth]{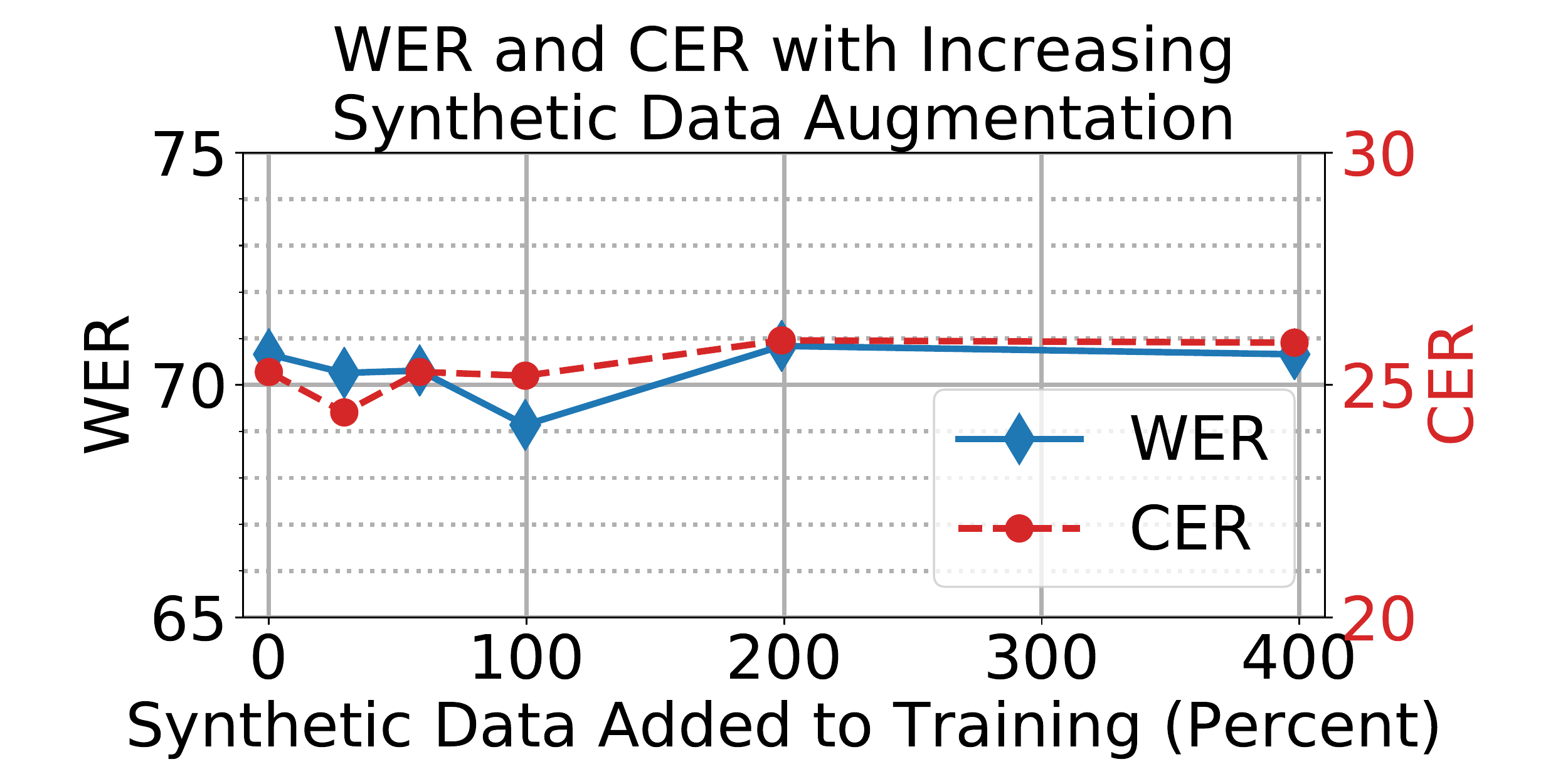}
\end{center}
\caption{Evaluation set WER and CER as synthetic (converted) speech data is added to ASR training set. Left-most point (0\%) corresponds to training the ASR system with only real data. Right-most point (400\%) shows the results when synthetic data four times the size of real data is added to the training set. Minor improvement is seen at up to 100\% synthetic data augmentation.}
\label{fig:wer}
\end{figure}

Fig. \ref{fig:wer} shows WER/CER values of the ASR system when increasing amounts of synthetic data is added to the training set. Results are obtained from an evaluation set of utterances from speakers not in the training set. Slight decrease in error rates is observed when synthetic data at up to 100\% of the real data size is added to the training set. This improvement vanishes when more synthetic data is used for augmentation.

\begin{figure}[t!]
\begin{center}
\includegraphics[width=0.7\columnwidth]{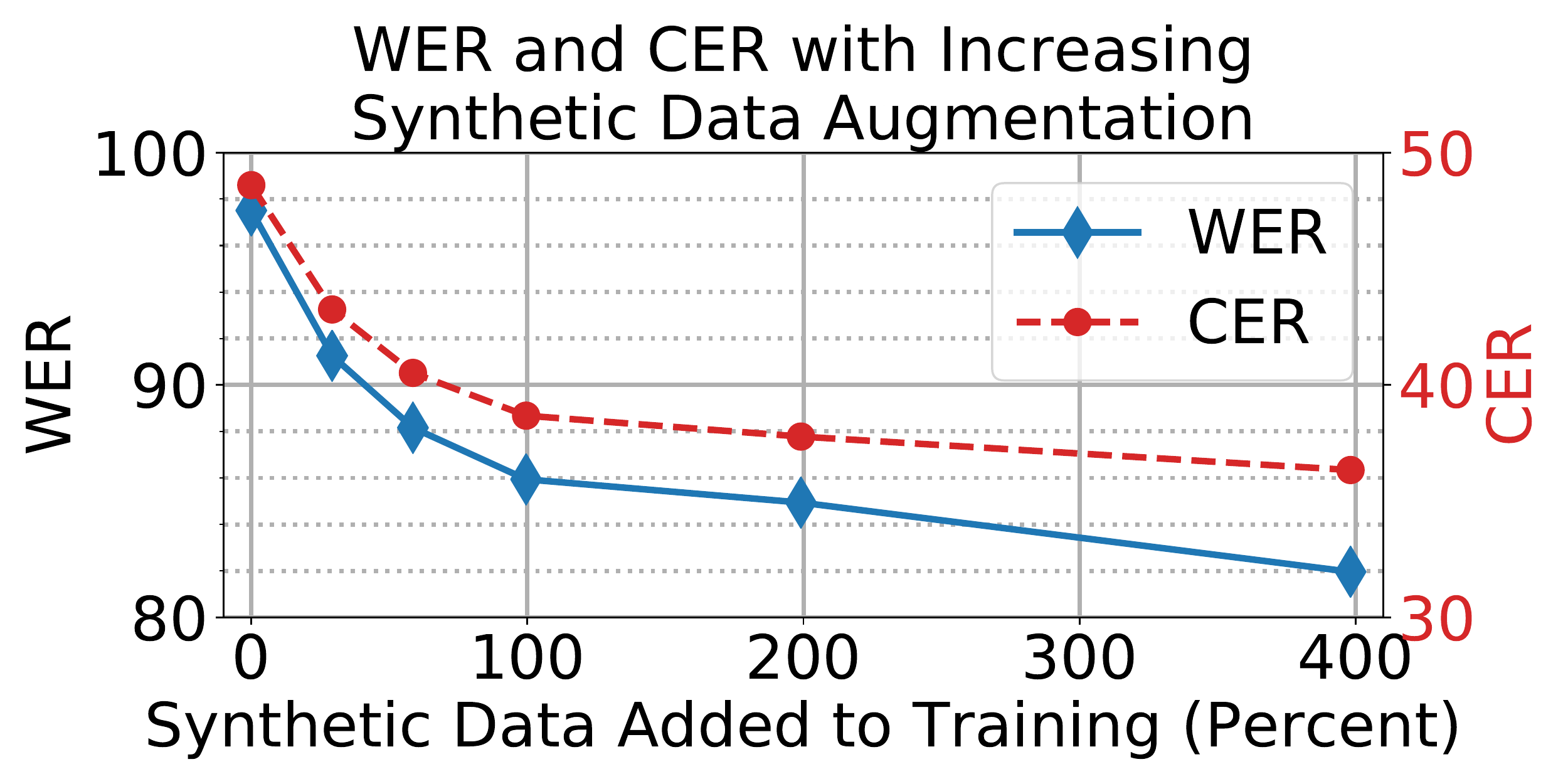}
\end{center}
\caption{WER and CER when evaluation set utterances are first converted to the style of training set speakers, and converted utterances are inferred by the ASR. WER and CER are lower for ASR systems trained with increasing amounts of synthetic data. However, error rates are significantly above the baseline in Fig. \ref{fig:wer}, where unconverted evaluation utterances are inferred by the ASR.}
\label{fig:wer_conv_val}
\end{figure}

Fig. \ref{fig:wer_conv_val} shows the error rates when evaluation set utterances are first converted to the style of training set speakers, and then passed through the ASR system. A decrease in error rates are seen as more synthetic data is added to the ASR training set. However, error rates are significantly higher than the case shown in Fig. \ref{fig:wer}, where evaluation utterances are directly inferred by the ASR system; making \textit{convert-then-infer} option undesirable for ASR.

\section{Discussion}
\label{sec:discussion}
In this paper, we report that a Cycle-GAN based voice converter model can generate audio files that are classified by four different automated SID models as the intended target speaker at up to 46\% top-1 accuracy. There is significant variation among SID models, with deep-learning based models having higher rates of intended target classification than i-vector based models. 

Additionally, we investigate if the high  imitation ability of the style, as rated by SID systems, can be used to improve error rates in ASR training. Our results demonstrate marginal improvement in WER/CER rates when the VC model is used to augment ASR training set. This latter result is in line with GAN-based augmentation methods seen in computer vision domain, and demonstrates further research is needed before VC models can be used to aid training in downstream tasks such as ASR. One future direction is to improve VC conversion quality further by using GAN architectures that can mimic styles in finer detail \cite{styleGAN}. Using a neural vocoder to re-build the raw audio waveforms from converted spectrograms to improve the quality of the conversions should also be investigated.

\bibliography{mybib}
\bibliographystyle{icml2019}



\end{document}